# An Infrared View of the Obscured AGN Environment in NGC 4945

G. Gaspar[1,2], R. J. Díaz[1,3], D. Mast[1,2], M. P. Agüero[1,2], M. Schirmer[4], G. Günthardt[1], and E. O. Schmidt[1,2,5]
[1] Observatorio Astronómico de Córdoba, Universidad Nacional de Córdoba, Laprida 854, X5000BGR, Córdoba, Argentina; gaiagaspar@gmail.com
[2] Consejo de Investigaciones Científicas y Técnicas de la República Argentina, Buenos Aires, Argentina
[3] Gemini Observatory, NSF's NOIRLab, USA
[4] Max-Planck-Institut für Astronomie (MPIA), Königstuhl 17, D-69117 Heidelberg, Germany
[5] Instituto de Astronomía Teórica y Experimental (IATE), CONICET-UNC, Córdoba, Argentina
Received 2021 November 11; revised 2022 February 18; accepted 2022 March 14; published 2022 April 25

## Abstract

NGC 4945 harbors one of the nearest active galactic nuclei (AGNs), which allows us to reach high spatial resolution with current observational facilities. The Seyfert 2 nucleus is deeply obscured by an edge-on disk with $A_V \sim 14$, requiring infrared observations to study circumnuclear structures and the interstellar medium. In this work, we present an imaging and long-slit spectroscopic study of the nuclear region with a spatial resolution of 6.5 pc, based on Flamingos-2 (F2) and T-ReCS data taken at the Gemini South observatory. We report subarcsecond photometric measurements of the nucleus in $J$, $H$, and $K_s$ filters, and at larger apertures. We do not detect nuclear variability. The nuclear spectra confirm that even in the $K$ band the AGN emission-line features are completely obscured by dust. We detect a circumnuclear disk in the $K$ band as well as in the mid-infrared $N$ and $Q_a$ bands, with a radial scale length of ∼120 pc. The disk shows knots mostly in a ring-like arrangement that has been previously detected with Paα observations from the Hubble Space Telescope, indicating that these are deeply embedded, massive young star clusters. We present here the spectrum of one of the brightest unresolved objects ($R < 5$ pc), which we identify as a super star cluster candidate with $M_{K_s} - 16.6 \pm 0.4$. For the circumnuclear region, a detailed rotation curve allows us to measure a nuclear mass of $M = (4.4 \pm 3) \times 10^6 \, M_\odot$ within a radius of ∼6.5 pc. We also report the detection of hot dust (∼1000 K) out to a nuclear distance of 80 pc measured along the semimajor axis.

*Unified Astronomy Thesaurus concepts:* Active galactic nuclei (16); Supermassive black holes (1663); Near infrared astronomy (1093); Galaxy nuclei (609)

## 1. Introduction

As deeper, higher-resolution data become available thanks to modern facilities such as the Atacama Large Millimeter/submillimeter Array (ALMA), Keck, and Gemini, among others, galactic centers have been discovered to be complex places. In the central kiloparsec of active galaxies can coexist superclusters of stars, large amounts of dust, and a supermassive black hole (SMBH) generating huge amounts of emissions and extreme motions in the circumnuclear material. Understanding the distribution, motion, and ultimate fate of the circumnuclear material has been proven to be essential in order to understand the mechanisms acting on these nuclei, which are proposed as one of the drivers of galactic evolution across cosmic times.

Several open issues regarding active galactic nuclei (AGNs) can be addressed by observations in weakly obscured IR and radio bands, with a spatial resolution better than a few tens of parsecs. AGNs are widely accepted to present a dust structure that is responsible for causing different types of observed emission depending on the line of sight (Blandford & Rees 1978; Antonucci 1993). Initially, this structure was proposed to be a homogeneous torus aligned with the equatorial plane of the SMBH, but the idea was quickly abandoned on the basis of energetic considerations (Krolik & Begelman 1988). Considerable progress was made with the inclusion of a clumpy torus (Nenkova et al. 2002, 2008). In recent years, more dynamic, complex, and extended torus models have been proposed that point to a clumpy distribution of outflowing dust that is no longer exclusively confined to the equatorial plane of the SMBH. Rather, it may have polar components, and their spatial extent may vary considerably among different nuclei (Ramos Almeida & Ricci 2017; Hönig & Kishimoto 2017; Carilli et al. 2019; Combes et al. 2019; Aalto et al. 2020; Garcia-Burillo et al. 2021). The characterization of these structures is of fundamental importance to understanding the Seyfert 1/Seyfert 2 dichotomy, and to constraining the fueling and accretion models of AGNs.

At larger scales, in the circumnuclear region inside the inner 100 pc, structures such as dust spirals, disks, rings, and bars have been observed and related to the funneling of gas toward the nucleus (Malkan et al. 1998; Martini et al. 2001; Díaz et al. 2003; Simões Lopes et al. 2007; Combes et al. 2013, 2014; García-Burillo et al. 2014; Schmidt et al. 2019). These structures appear more frequently than expected in active galaxies (Pogge & Martini 2002; Martini et al. 2003; Agüero et al. 2016). Their appearance in non-active galaxies may be ascribed to the difference in dynamical timescales of the involved phenomena (likely $10^4$–$10^5$ yr, e.g., Schawinski et al. 2015; Novak et al. 2011), which are much shorter than the dynamical timescale of galaxies (of the order of $10^8$ yr). Therefore not every galaxy with these features will currently be in a phase with nuclear activity, yet they seem to be responsible for the transport of material from scales of kiloparsecs down to 100 pc. Another fueling mechanism could be active in the last 100 pc to the active nucleus. Some authors have explored theoretical mechanisms that could cause departures from axisymmetry in the nuclear galaxy potential, which produce torques over the circumnuclear material, (Emsellem et al. 2015; Capelo & Dotti 2017; Kim & Elmegreen 2017), but the observational evidence is still scarce. Multiwavelength studies of the neighborhood of the SMBH are thus essential in order to elucidate the conditions under which the nuclear activity is triggered.

The weakly obscured $K$ band is particularly useful for investigating this topic further. Emission lines from molecular and ionized gas can be used to probe the excitation state of the gas, its turbulent motion, and rotational bulk flows. Furthermore, the dust heated by the AGN's accretion disk emits strongly in the $K$ band, such that its temperature and extent can be determined (Ferruit et al. 2004; Burtscher et al. 2015; Durré & Mould 2018; Gaspar et al. 2019; Gravity Collaboration et al. 2020). Owing to the low obscuration by dust (Mathis 1990), one can map the embedded stellar structures.

In this paper, we present $K$-band images and spectra of NGC 4945 obtained with F2 at Gemini South (Eikenberry et al. 2012; Díaz et al. 2013), along with complementary mid-infrared (MIR) images from T-ReCS (Gemini South, Telesco et al. 1998).

NGC 4945 is a nearby galaxy (3.72 Mpc, Tully et al. 2013) that harbors an obscured Seyfert 2 nucleus (Madejski et al. 2000; Goulding & Alexander 2009). The high nuclear obscuration has made it difficult to determine the true nature of its nuclear emission. There is no signature of an AGN in optical bands (Koornneef 1993; Spoon et al. 2000), yet X-ray emission (e.g., Madejski et al. 2000; Schurch et al. 2002; Rosenthal et al. 2020) and the presence of the [Ne V] 14.32 $\mu$m (97.1 eV) emission line (Goulding & Alexander 2009) reveal its presence. Furthermore, the mass of the compact nuclear object has been measured by Greenhill et al. (1997) via water maser kinematics inside a radius of 0.7 pc, and the location of the kinematic center of the maser coincides with the peak of the $K$-band emission within the astrometric uncertainties (Marconi et al. 2000). A circumnuclear ring of star formation was first proposed by Moorwood et al. (1996) on the basis of Br$\gamma$ 2.16 $\mu$m emission and mapped in Pa$\alpha$ 1.87 $\mu$m by Marconi et al. (2000). In the radio regime, at 93 GHz free–free continuum, the ring is not observed, but several superclusters of star formation are unveiled by ALMA (Emig et al. 2020).

## 2. Observations and Data Reduction

### 2.1. Near-IR (NIR) Spectroscopy

$K$-band long-slit spectra were taken with F2 (Eikenberry et al. 2008; Gomez et al. 2012) at Gemini South in queue mode (Program ID: GS-2019A-Q-121). We used the $K_{\rm long}$ filter that covers a wider wavelength range of 1.9–2.5 $\mu$m than the older $K_{\rm s}$ filter, thus reaching the start of the CO band at 2.4 $\mu$m. We also used the 3 pixel wide slit (0.″54 width) and the R3K grism, yielding a spectral resolution of $R \sim 2100$ in the center of the band. The corresponding dispersion is $\sim$3.5 Å pixel$^{-1}$. The spatial resolution is 0.″18 pixel$^{-1}$. The position angle (PA) of the slit was set to 43°, aligned with the major axis of the galaxy.

NGC 4945 has a $K_{\rm s}$ Vega magnitude of 13.55 inside 2.″5 as measured by the Two Micron All Sky Survey (2MASS).[6] With seven spectroscopic exposures of 300 s on source, we reach a combined signal-to-noise ratio (S/N) $\sim$ 150 in the center of the band. We adopted an ABCD nod configuration where A and B were target nods dithered by 10″, and C and D were sky positions 5′ from the target, with the same dither offset. The average seeing during the observation was 0.″5, yielding a spatial resolution of 9 pc (1″ = 18 pc) at the distance of NGC 4945.

For the reduction of the spectra we used the hybrid procedure described in Gaspar et al. (2019). The Gemini PyRAF tasks of the F2 data reduction recipe were used for the dark subtraction, flat-fielding, sky subtraction, and combination of 2D spectra. Wavelength calibration and telluric standard correction were performed with standard IRAF[7] tasks. In particular, the telluric correction was improved with respect to the F2 data reduction recipe: we modeled a Lorentzian profile to the telluric's intrinsic Br$\gamma$ absorption line at 2.1654 $\mu$m using splot, and subtracted the fit from the data. The result was divided by a stellar continuum template built with the IRAF task mk1dspec. The resulting atmospheric extinction spectrum was then used to correct the spectrum of NGC 4945.

### 2.2. NIR Imaging

The $JHK_{\rm s}$ F2 images were also taken through GS-2019A-Q-121. We took 12 dithered exposures per filter, six on target and six on sky, with an exposure time of 2 s. The images were reduced using standard basic calibration and sky subtraction using THELI v3.0 (Schirmer 2013; Erben et al. 2005). The astrometric calibration was performed against Gaia DR2 (Gaia Collaboration et al. 2018). The photometric calibration was performed against 2MASS point sources, and translated from Vega to AB mag using the conversion factors of Pons et al. (2019). THELI v3.0 scales the coadded images in photoelectrons s$^{-1}$, for which we obtain the following AB mag photometric zero-points:

$$ZP_J = 24.38 + 0.29(J - H) \quad (\sigma = 0.01) \tag{1}$$

$$ZP_H = 25.91 - 0.30(J - H) \quad (\sigma = 0.01) \tag{2}$$

$$ZP_{K_{\rm s}} = 26.09 - 0.45(J - K_{\rm s}) \quad (\sigma = 0.02). \tag{3}$$

Section 3.1 reports the brightness measurements in different apertures using these zero-points.

### 2.3. MIR Imaging

For the MIR imaging we used archival T-ReCS (Telesco et al. 1998) data taken at Gemini South on 2006 March 17 and April 17 (Program ID: GS-2006A-Q-62, PI James Radomski). Images were taken in filters $Si$-2 (8.7 $\mu$m), $N$ (10.36 $\mu$m), and $Q_{\rm a}$ (18.3 $\mu$m), with total on-source exposure times of 261, 304, and 261 s, respectively. Again, the data were processed with THELI v3.0, using the standard MIR chop–nod sky subtraction technique. The plate scale of the T-ReCS data is 0.″089 pixel$^{-1}$, with an average spatial resolution of $\approx$0.″35 ($\approx$6 pc). The T-ReCS data were astrometrically registered to the F2 data, using common bright emission knots visible in the MIR data and the F2 $K_{\rm s}$-band images.

---

[6] https://irsa.ipac.caltech.edu/applications/2MASS/LGA/atlas

[7] http://ast.noao.edu/data/software

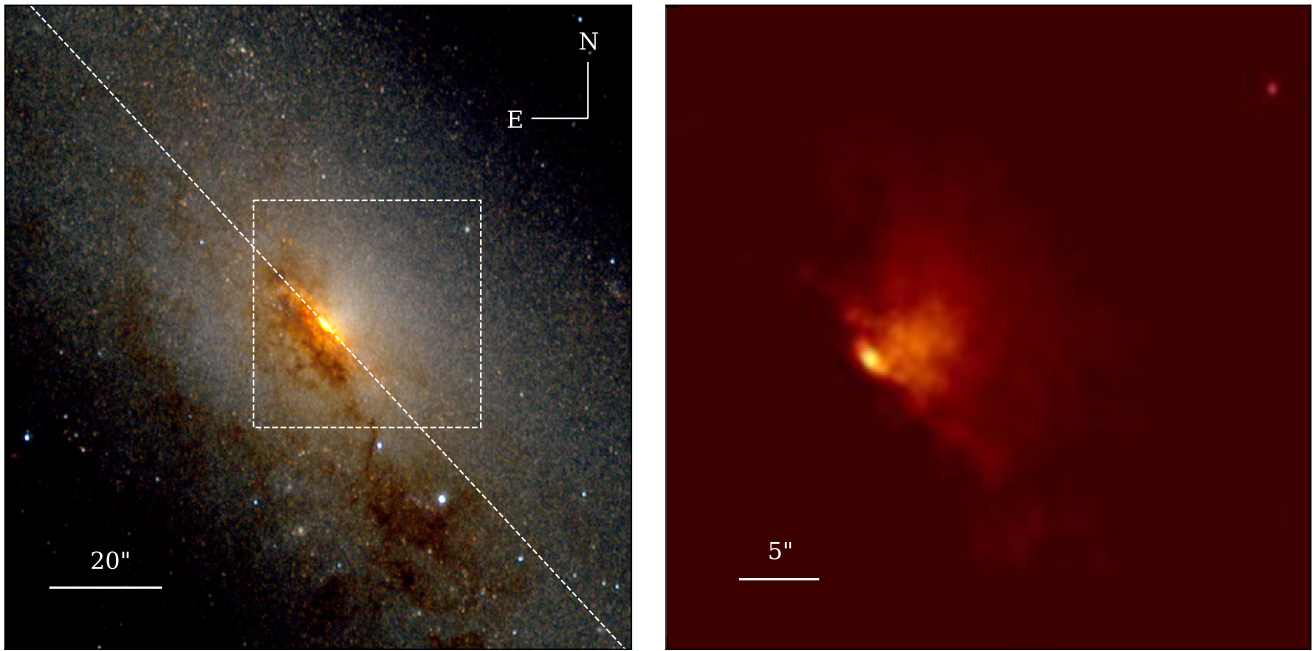

**Figure 1.** F2 images of the NGC 4945 nucleus. Left panel: RGB color composite of the $JHK_s$ images, with $J$, $H$, and $K_s$ rendered blue, green, and red, respectively. The white dotted line shows the PA and the width of the long slit used, and the white square indicates the area enlarged in the right panel. Note how the prominent dust lanes approach the nucleus, forming a tightly wound structure in the inner 200 pc (for this galaxy $1'' = 18$ pc). Right panel: zoomed view of the nuclear region in the $K_s$ band (seeing $\sim 0\rlap{.}''5$). The contrast was chosen to show material being illuminated above the equatorial plane of the system, plausibly corresponding to the AGN's ionization cone.

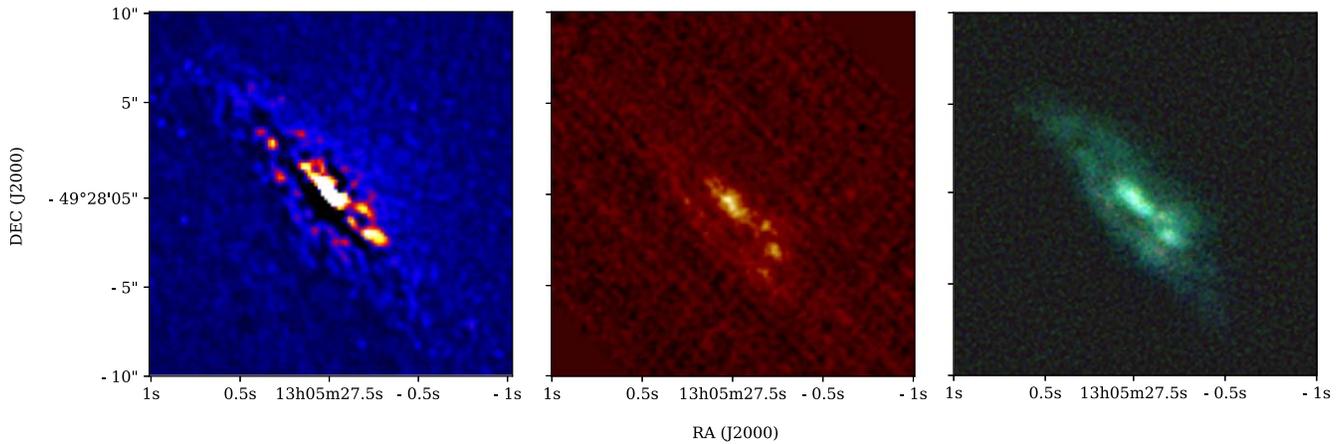

**Figure 2.** Left panel: unsharp mask image of the circumnuclear region of NGC 4945 from the F2 $K_s$-band image, showing a circumnuclear disk structure. Middle panel: $Q_a$-band image from T-ReCS. An arc of compact MIR emission knots is detected SW of the nucleus. Right panel: MIR pseudo-color composition of the T-ReCS data, $Si$-2 (8.7 $\mu$m) = blue, $N$ (10.36 $\mu$m) = green, and $Q_a$ (18.3 $\mu$m) = red, depicting a circumnuclear spiral structure dominated by emission in the $N$ band. The three images display the same field of view.

## 3. Results

### 3.1. Circumnuclear Structures

As described above, the nucleus of this galaxy has been widely studied in different wavelengths. In particular, the nuclear activity is partially masked by the intense starburst activity within the central $\sim$100–200 pc. In this section, we study this region with a combination of NIR and MIR imaging, and NIR spectroscopy. The NIR data provide sufficient dust penetration, revealing both the embedded stellar structures and the presence of hot dust, while the MIR images locate the cold dust. Both data sets have a spatial resolution of $0\rlap{.}''5$ or better, complementing existing Hubble Space Telescope (HST) optical and ALMA radio observations.

In Figure 1 we present two displays of F2 images. The left panel shows an RGB composition of the $JHK_s$ images of $2' \times 2'$ size. Prominent dust lanes cover the path from the kiloparsec scale down to a scale of 100 pc, where the dust lanes seem to coil around the nuclear region inside a radius of 200 pc. This structure is coincident with the widely studied starburst that resides in this nucleus, but the high inclination of the galaxy hides the true morphology of this inner dust structure in the $K_s$ band. The reddest image we have, taken with T-ReCS in the $Q_a$ band, shows discrete knots, which form part of a nuclear disk with fuzzy nuclear spiral structure as detected in the $Si$-2 and $N$ bands (Figure 2). The radial range of the MIR spiral structure (30–70 pc) is the same as the structure that appears as a circumnuclear ring in our $K_s$ images (Figure 2) and seems in accordance with the star-forming ring mapped in ionized gas by Moorwood et al. (1996). More recently, Emig et al. (2020) presented ALMA observations in 93 GHz continuum (star formation tracer) and in 350 GHz emission (dust tracer). The images they presented show 29 candidates for super star clusters in 93 GHz, which they identified as discrete sources that are not coincident with the Paα knots that form the ring described by Moorwood et al. (1996). Furthermore, no Paα emission is detected by HST inside the dense dust structure seen in our Figure 1, probably due to the high obscuration in the region. All the 29 sources detected by ALMA are located inside the dust structure of ∼200 pc radius seen in the $K_s$ image.

In the right panel of Figure 1 a zoom of the innermost region in the $K_s$ band is presented; the position of this region is marked in the left panel with a white square of ∼0″.5 side. A cone-shaped structure is observed with an extent of approximately 1″.5 or 270 pc perpendicular to the galaxy plane. The structure is found to be patchy but the $K_s$ image does not provide information about the nature of this emission. Other authors have mapped a global cone shape at much larger scales; for example, Moorwood et al. (1996) presented Hα-, H$_2$-, J-, and Z-band images where the cone is seen; they attribute the emission to starlight seen through extinction and rule out any AGN contribution to the cone on energetic grounds based on the J-band flux calculations and the lack of [O III] emission. In other work, Venturi et al. (2017) use MUSE observations of NGC 4945 taken through the MAGNUM survey (Mingozzi et al. 2019) to map this ionization region in [N II]. Venturi et al. (2017) describe a biconical structure where the NW lobe reaches 1.8 kpc and the SE lobe is smaller and weaker because it is almost completely extinguished by dust (the SE lobe is not visible in the $K_s$ F2 image). The velocity map of the NW lobe shows approaching velocities at the edges of the cone and receding ones along its axis, while the opposite behavior is found in the SE cone in consistency with a biconical structure. On this basis, the authors propose that the [N II] NW lobe follows a hollow cone with a sufficient aperture such that the farther part of the cone intercepts the plane of the sky. The small-scale cone that we detect in the K band is connected to the larger-scale structure; its opening angle (≈100°) is similar to the opening angle at the vertex of the outflow system mapped in [N II] emission by Venturi et al. (2017, their Figure 2, panel (A)).

In Table 1 we report J-, H-, and $K_s$-band magnitudes for different apertures centered on the brightness peak. The K-band images show the nucleus as a compact source at the vertex of the cone structure, with a brightness of 0.014 times the brightness of the circumnuclear region in the central 20″ (Table 1). The 4″ magnitude is, within the uncertainties, the same as reported by Skrutskie et al. (2006) from 2MASS measurements. Other measurements (Marconi et al. 2000; Moorwood & Glass 1984) lie within ∼0.2 mag and can also be considered to lie within the uncertainties, taking into account differences in filters and calibration. We conclude that the emerging radiation at NIR wavelengths is highly processed because we do not detect significant photometric variability directly from the AGN, when comparing with previous photometric measurements.

In Figure 3 we present the nuclear spectrum from the central spatial pixel (3.25 pc width), to minimize contamination from the circumnuclear region. The nuclear spectrum displays

**Table 1**
J, H, and $K_s$ Magnitudes Measured in Different Apertures Centered on the Nucleus

| Aperture radius (arcsec) | J (mag) (±0.05) | H (mag) (±0.01) | $K_s$ (mag) (±0.02) | Literature K-band data (mag) |
|---|---|---|---|---|
| 0.25 | 18.09 | 14.85 | 12.82 | |
| 0.5  | 16.8  | 13.55 | 11.60 | |
| 1    | 15.68 | 12.5  | 10.64 | |
| 2    | 14.63 | 11.64 | 9.86  | |
| 3    | 13.94 | 11.09 | 9.38  | 9.14[a], 9.34[b] |
| 4    | 13.46 | 10.7  | 9.06  | (8.991 ± 0.007)[c] |
| 6    | 12.81 | 10.19 | 8.67  | |
| 9    | 12.15 | 9.65  | 8.26  | 8.12[b] |
| 10   | 11.98 | 9.51  | 8.16  | |

**Notes.** For comparison we compiled $K_s$ magnitudes from the literature in the last column.
[a] F222M NICMOS magnitude from Marconi et al. (2000).
[b] K band from Moorwood & Glass (1984).
[c] 2MASS $K_s$ band from Skrutskie et al. (2006).

several emission lines (Table 2), among others those from molecular hydrogen. H I and He I emission lines from ionized hydrogen and helium are also present, as well as parts of the CO absorption bands at the long-wavelength end of the spectrum. Absorption lines from Mg, Al, Na, and Ca are also detected. The gray shaded region in Figure 3 corresponds to the spectral region around 2 μm affected by strong atmospheric absorption, increasing the noise level after telluric correction.

As mentioned in Section 1, optical spectra of NGC 4945 do not reveal AGN features because of heavy dust obscuration (Spoon et al. 2000; Koornneef 1993). Observations in the far-IR (Brock et al. 1988), MIR (Goulding & Alexander 2009), and X-rays (Madejski et al. 2000) were needed to determine the Seyfert nature of this nucleus beyond doubt (Ackermann et al. 2012; Peng et al. 2019; Rosenthal et al. 2020). Moreover, our deep K-band spectra do not reveal features conclusive of an AGN because coronal lines are absent, and the FWHM of the emission lines is typical for star-forming regions. The AGN is so deeply buried that its light is extinguished even at these wavelengths where the extinction is reduced to 10% of that observed in the optical V band (Mathis 1990). This deep obscuration is consistent with the considerable drop in Brγ EW at the galactic core, as shown in Section 3.3.

### 3.2. Spatial Profiles

Spatial extractions have been performed in three regions of the K-band spectrum: a 600 Å wide extraction at λ = 2.038 μm, a 28 Å wide extraction centered on the H$_2$ emission line at λ = 2.121 μm, and a 28 Å wide extraction with center in the Brγ line at λ = 2.165 μm. All three regions are marked with blue rectangles in Figure 3. The spatial profiles are presented in Figure 4. In this section—and throughout this paper—negative radii correspond to the SW direction and positive radii to the NE direction.

In the upper panel of Figure 4 the 2.038 μm continuum extraction is presented. The profile is well modeled by the sum of a nuclear source (represented by a Gaussian model) and a disk, represented by a Sérsic profile with Sérsic index n = 1 and a scale length of (117 ± 7) pc. The region around −55 pc

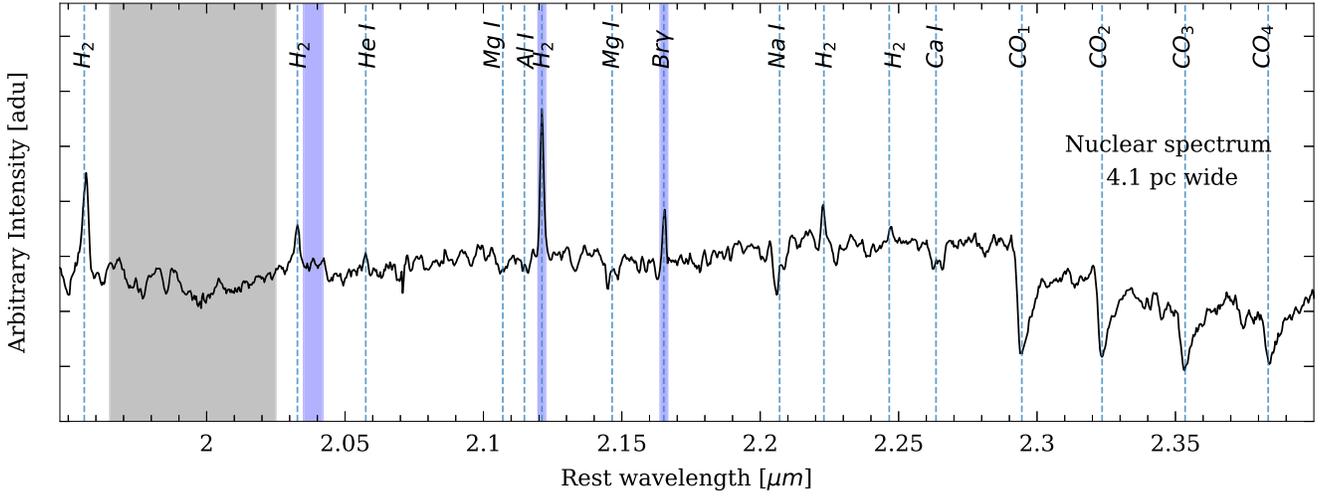

**Figure 3.** *K*-band spectrum of the NGC 4945 nucleus, extracted from a 1 × 1 pixel (0."18 × 0."18) wide spatial region centered on the nucleus. The shaded blue bands identify the regions where the spatial profiles were extracted (see text), while the gray band identifies a region of high telluric absorption. The nuclear emission is characterized by strong molecular hydrogen lines and the Brγ recombination line. We use them to characterize the state of the different gas phases as well as the dynamics of the material around the central SMBH.

**Table 2**
Emission and Absorption Lines Present in the Nuclear Spectrum of NGC 4945

| Ion/Molecule | Wavelength (μm) | EW (Å) |
|---|---|---|
| $H_2$ | ... | 9.7 |
| $H_2$ 1–0S(2) | 2.0332 | 3.7 |
| He I | 2.058 | 1.2 |
| $H_2$ 1–0S(1) | 2.1211 | 8.7 |
| H I (Brγ) | 2.1654 | 3.2 |
| $H_2$ 1–0S(0) | 2.2230 | 2.3 |
| $H_2$ 2–1S(1) | 2.2470 | 1.2 |
| Al I (abs) | 2.1099, 2.1169 | 1.9 |
| Mg I (abs) | 2.145, 2.1481 | 3.9 |
| Mg I (abs) | 2.1066, 2.1067 | 2.5 |
| Na I (abs) | 2.2062, 2.2089 | 4.4 |
| Ca I (abs) | 2.2626, 2.2656 | 0.2 |
| $CO_1$ (abs) | 2.2935 | 14.48 |
| $CO_2$ (abs) | 2.3227 | 16.17 |
| $CO_3$ (abs) | 2.3525 | 18.33 |
| $CO_4$ (abs) | 2.3829 | 17.42 |

**Note.** The first column identifies the spectral features, marked with (abs) in the case of absorption lines. The second column lists the rest-frame wavelength. The Mg I, Al I, Na I, and Ca I absorption features are doublets. For Mg I, the doublet wavelengths were not found in the literature and we report the measured values. The last column provides the equivalent line width as measured in the nuclear spectrum (Figure 3).

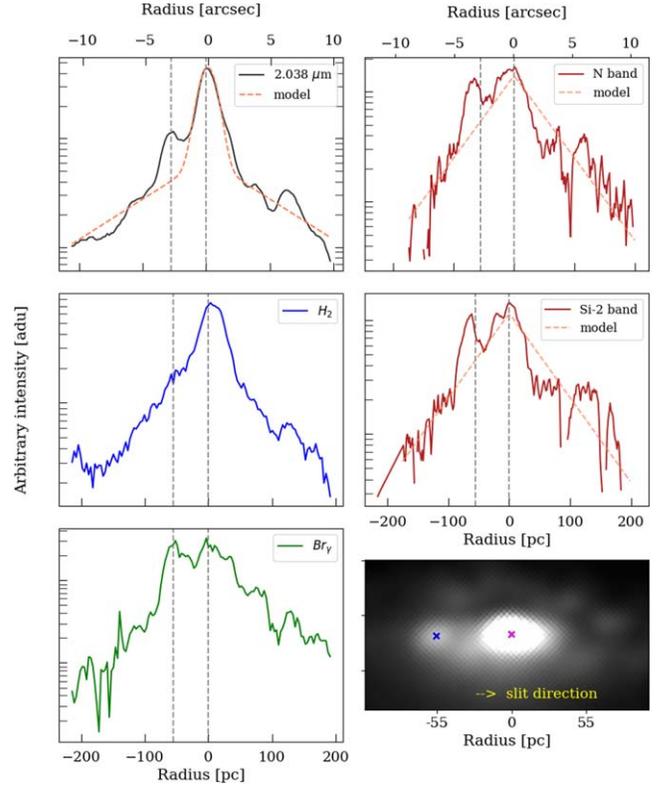

**Figure 4.** Spatial profiles for the nuclear region of NGC 4945. Upper left panel: in black the stellar continuum extracted at 2.038 μm. The red dashed curve corresponds to a spatial component model consisting of a disk and a Gaussian that accounts for the nuclear source. Middle left panel: the intensity profile extracted in $H_2$ 2.121 μm. Bottom left panel: the intensity profile extracted in Brγ 2.165 μm. The dashed vertical line at −55 pc corresponds to the position of the source SWK. Upper and middle right panels: spatial profiles for NGC 4945 in the *N* band (centered at 10.36 μm) and the *Si*-2 band (centered at 8.7 μm) respectively. Both profiles have been modeled by a disk component. Bottom right panel: $K_s$ image of the nuclear region; the magenta cross marks the position of the active nucleus while the blue cross marks the position of SWK region's core, an SSC candidate at 55 pc in the SW direction.

that exceeds the model corresponds to a source visible in the $K_s$-band image (bottom right panel of Figure 4). This source (hereafter SWK) does not correspond to any of the sources found by Emig et al. (2020). The location of SWK also appears prominent in the Brγ profile, presenting a flux comparable to that from the nucleus itself. SWK also displays a small but appreciable excess in the $H_2$ profile (middle left panel of Figure 4).

We extracted spatial profiles along the major axis of the nuclear disk on the T-ReCS images in *Si*-2 (8.7 μm) and *N* (10.36 μm). These extractions have the same width and PA as the F2 long slit (0."54 width, PA = 43°). The obtained profiles are presented in the upper and middle right panels of Figure 4. We modeled both profiles with an exponential disk of 117 pc with an uncertainty of 5 pc for the *N* band and 4 pc for the *Si* band. The scale length of the MIR disk is coincident with that measured in the $K_s$ band. The SWK knot appears as the most important feature outside the nucleus itself, both in the Brγ profiles and at the MIR wavelengths. This reveals that this star-forming complex is also deeply embedded in warm dust.

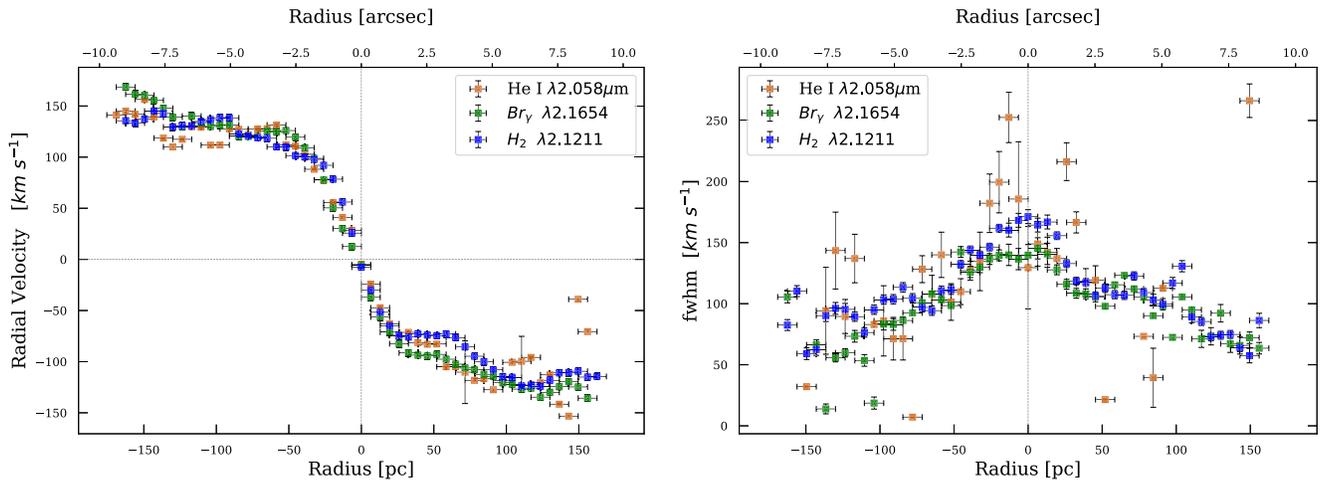

**Figure 5.** Left panel: radial velocity curves for two hydrogen and one helium lines. The zero velocity was calculated separately for each line and corresponds to the average of the five inner independent velocity measurements, excluding the nucleus. Right panel: FWHM measurements for the same three lines.

In the bottom right panel of Figure 4 we present the $K_s$ image of the circumnuclear disk, showing the nucleus and the SWK source; the distance between them is 55 pc.

### 3.3. Physical State of Gas

In Figure 5 we present radial velocity curves and FWHM measurements for two hydrogen emission lines, $H_2$ 2.12 μm and Brγ 2.16 μm, and for He I 2.058 μm. The three radial velocity curves present the characteristic symmetric shape of rotation curves of the central portion of galactic disks. Thermal turbulence and shocks affect the ionized gas ($T \geqslant 10^4$ K, as measured from [Fe II] lines, e.g., Rodríguez-Ardila et al. 2004), which has larger noncircular motions and is distributed in a wider disk than relatively cold molecular gas ($T \leqslant 5000$ K, see Mazzalay et al. 2013; Riffel et al. 2021). In contrast, the molecular gas has been observed to be less affected by turbulence and to lie in a thinner disk in the equatorial plane of the system (Storchi-Bergmann et al. 2009), and hence it is expected to present a steeper slope in its radial velocity curve. However, in the case of NGC 4945, no significant difference is observed between the radial velocity curves of the molecular and ionized gas. Furthermore, the FWHM curves for the same lines (Figure 5, right panel) do not show conclusive differences between the $H_2$ curve (blue) and the Brγ curve (green). The last one is somehow puzzling due to a "plateau" that appears in the nuclear region around ∼140 km s$^{-1}$. A drop in velocity dispersion like this has been detected in H recombination lines at optical wavelengths for several star-forming galactic nuclei and it has been explained as a fall in the velocity dispersion of the star-forming gas due to a strong accumulation of gas in a dissipative disk (e.g., Falcón-Barroso et al. 2006). In addition, the extinction at the nucleus might be so high that the gas with higher velocity dispersion associated with the active nucleus is not detected. This is consistent with what we see in Figure 7, where the Brγ EW rises steadily toward the center with a sudden drop in the most central spectral slice. The He I FWHM curve (Figure 5, pink) shows higher values in the nucleus, as expected, but the errors are large and hence no conclusive trends are found.

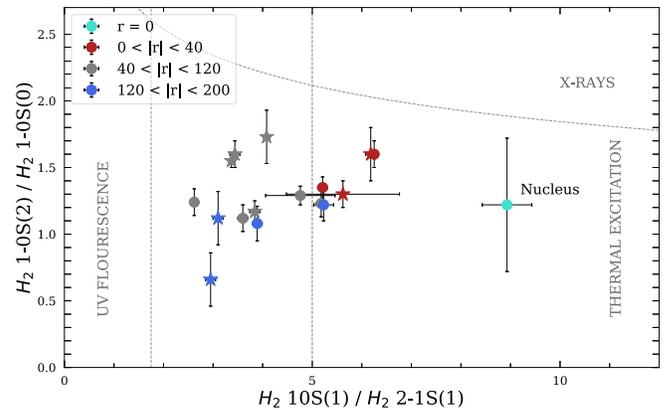

**Figure 6.** Line ratio diagram for the molecular hydrogen lines. The dotted vertical lines separate the regions where different excitation mechanisms dominate (see labels). The region ∼2–5 on the *x*-axis corresponds to a mixture of UV fluorescence and thermal excitation. The color code distinguishes between different intervals of radius, and directions along the slit are differentiated by symbol: circles correspond to the NE direction and stars to the SW.

The uncertainties of the radial velocities were estimated using known sky lines along the slit. This resulted in an upper limit for the uncertainties of 3 km s$^{-1}$, in accordance with the values obtained by Gaspar et al. (2012) for the uncertainties in the measurement of the velocity for a single line of high S/N with F2 spectra. The uncertainties for the FWHM in the Gaussian fits were estimated for bins of ∼40 pc width by measuring each line 10 times, changing the starting parameters for the Gaussian fit, and then calculating the standard deviation of all the measurements. The error bars in the spatial direction correspond to the size of the extraction aperture for each spectrum (6.25 pc).

In Figure 6 we present a diagnostic diagram that uses molecular hydrogen lines to determine the dominant excitation mechanism of the lines involved. For a detailed description of the construction of the diagram and the theory behind it see, for example, Reunanen et al. (2002), Ramos Almeida et al. (2009), Falcón-Barroso et al. (2014), and Günthardt et al. (2015).

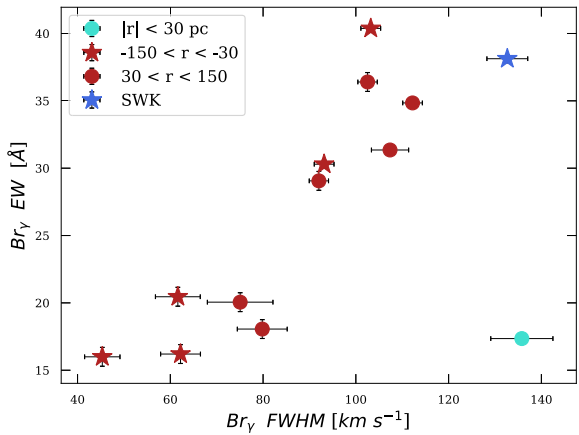

**Figure 7.** FWHM vs. equivalent width for the Brγ hydrogen recombination line. The color code distinguishes between different radius intervals, and directions along the slit are differentiated by symbol: circles correspond to the NE direction and stars to the SW.

Briefly, the diagram uses two ratios involving four hydrogen lines of the K band: 1–0S(2) 2.03 μm, 1–0S(0) 2.22 μm, 1–0S(1) 2.12 μm, and 2–1S(1) 2.25 μm. These ratios are indicators of the mechanism that dominates the emission of the lines, which can be thermal (shocks or UV/X-ray radiation) or nonthermal (UV fluorescence). Also, both cases can happen simultaneously. The region between ∼2 and 5 on the X-axis indicates that the lines are excited by a mixture of UV fluorescence and thermal excitation by shocks. Other authors have presented similar diagrams for NGC 4945 (e.g., Moorwood & Oliva 1990); the novelty of the one presented here resides in the high spatial resolution of our K-band spectra that allows us to place the line ratios for several radii separated by 20 pc from the nucleus to ∼200 pc in both slit directions. The turquoise point corresponds to the line ratios measured from the nuclear spectrum. The nucleus of NGC 4945 falls in the thermal excitation regime as well as various other regions of (mostly) small radius (red symbols, $|r| < 40$ pc). For larger radii (blue and gray symbols, 40 pc $< |r| <$ 200 pc), the molecular hydrogen begins to enter in the zone of mixed excitation mechanism, showing that UV fluorescence begins to gain importance in more external regions. This effect may arise due to the lowering of the dust extinction at the external radii of the dusty structure detected with the red excess in the spectra (∼100 pc). If this structure harbors a strong starburst it is expected that a large number of UV photons from the young stars are available. As the structure becomes less dense with increasing radius, those photons have a longer free path before being absorbed by the dust, and hence are capable of exciting the molecular hydrogen. As expected, the region around SWK marked with gray stars in the diagram shows higher excitation than its radio counterpart marked with gray circles.

The ionized gas, probed by the FWHM versus EW for Brγ (Figure 7) shows a rising linear behavior toward the center, indicating that the number of ionizing photons rises toward the central source. The sudden and significant drop in the nuclear values in relation to the rest of the circumnuclear region indicates that some other mechanism is decreasing the EW of the line, probably the large amounts of dust present in the nuclei. The drops in EW might also be due to the higher continuum emission in the nucleus; however, the line intensity spatial profile (Figure 4, bottom left panel) shows that Brγ emission does not rise significantly at the AGN position. In fact, it is comparable to the SWK Brγ emission.

### 3.4. The SWK Source

We have identified in our F2 spectra and images a source visible in the K band but not present in Emig et al. (2020), which we have named SWK (southwest knot). This source can be identified in the $K_s$ image presented in Figure 4 as a secondary source 55 pc southwest of the nucleus.

As modeled by an ellipse coincident with the more external closed contour around the source, the position of SWK is R. A. = 13:05:27.23, decl. = −49:28:07.25. The semimajor and semiminor axes lengths are 0″.514 and 0″.352 respectively, or equivalently 9.25 pc and 6.34 pc. The brightness is dominated by an unresolved core with FWHM similar to the seeing (∼0″.5 = 9 pc), therefore the central object must have a diameter of less than 5 pc in order to still appear unresolved under the observing conditions and detector spatial sampling. This size limit is consistent with the size of the young star cluster candidates observed in NGC 4945 by Emig et al. (2020) at radio wavelengths. We measured the unresolved core brightness with an aperture of radius 0″.5 (∼9 pc) to be $(12.85 \pm 0.02)$ mag. Considering the distance to NGC 4945, this represents an uncorrected absolute magnitude $M_{K_s} = -15.00 \pm 0.02$. Moorwood & Oliva (1988) and Marconi et al. (2000) estimate $A_V \sim 14 \pm 3$, which, assuming $R_V = 3.1$ (Cardelli et al. 1989), means a dereddened $M_{K_s} = -16.6 \pm 0.4$, where the error is dominated by the uncertainty in $A_V$. The object has a K-band brightness typical of super star clusters (SSCs) found in luminous infrared galaxies and those with a high star formation rate (see Randriamanakoto et al. 2015, their Figure 1).

The overall $K_s$ brightness of the region, measured in an aperture of 0″.9 radius, is $(11.89 \pm 0.02)$ mag. The extinction of this region is high enough to extinguish the Paα emission of the source (Emig et al. 2020) but, in contrast, the Brγ emission is prominent in our profile (see the bottom left panel of Figure 4). The T-ReCS profiles in the N band (centered at 10.36 μm) and Si-2 band (centered at 8.7 μm) also show significant emission around SWK, reinforcing the scenario where this is a deeply dust-embedded star formation region populated by one or more SSCs.

In Figure 8 we present the spectrum for the SWK source. The relatively high intensity of Brγ with respect to the $H_2$ lines is evident in comparison with the nuclear spectrum presented in Figure 3. The intensity of the He I line is also higher than that observed in the nuclear spectrum.

### 3.5. A Nuclear Gas Reservoir

As mentioned in the previous section, the molecular hydrogen is a relatively cold phase, and as shown in Figure 5 it is less turbulent than the ionized phase (as represented by the He I line). Due to this, the molecular hydrogen is expected to be confined to a disk at low galactic latitude. This makes the $H_2$ lines good tracers of the gravitational potential in the circumnuclear regions and hence they can be used to estimate the mass of the material that is causing the potential. Under this hypothesis, the $H_2$ radial velocity curve (Figure 5) can be used to measure the mass of the SMBH and the circumnuclear material inside a certain radius. In particular, for NGC 4945 we have mapped the velocity curve with a resolution of 6.5 pc.

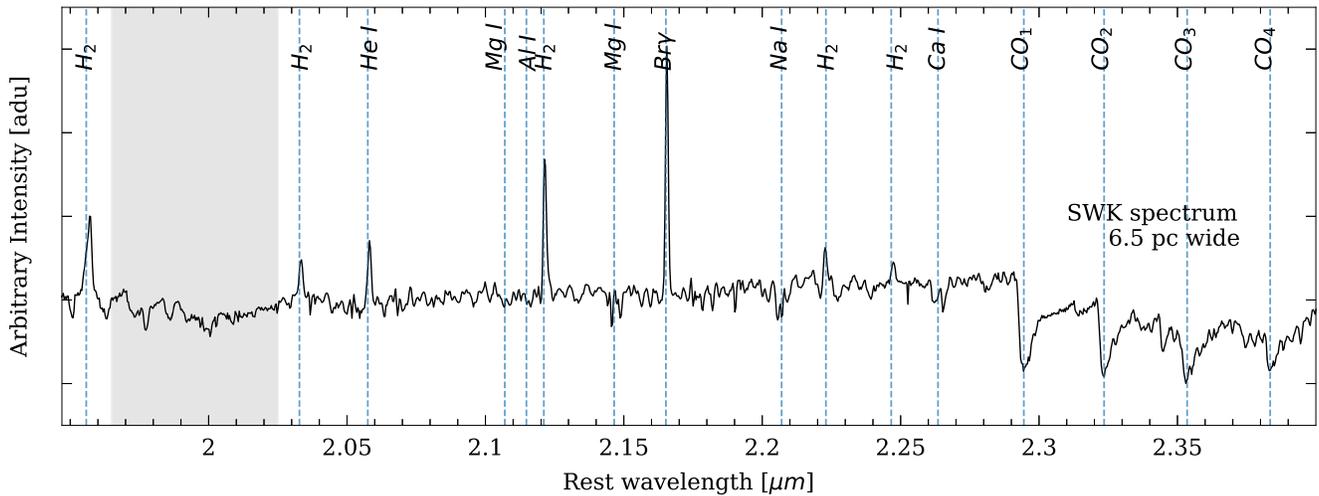

**Figure 8.** $K$-band 6.5 pc wide spectrum of the SWK source. Note the increased Br$\gamma$/H$_2$ ratio in comparison with the nuclear spectrum presented in Figure 3, indicating a higher star formation rate in the F2 source.

Then, the curve allowed us to estimate the mass enclosed inside a radius of $6.5 \pm 3.25$ pc using the Keplerian approximation for the mass enclosed in a radius $R$ [pc] that produces a velocity amplitude $V$ [km s$^{-1}$] in the gas: $M = 233 V^2 R / \sin(i)$, where the resulting mass is in solar masses and $i$ is the inclination of the system. Applying this expression, we measured a mass for the material enclosed inside $6.5 \pm 3.25$ pc of $M/M_\odot = (4.8 \pm 2.9) \times 10^6 \sin(i)$.

We have estimated the inclination of the system by modeling the nuclear disk as seen in the $N$ band (Figure 2) using ellipses and obtained a value of $i = (67.5 \pm 2.5)^\circ$. This yields $M/M_\odot = (4.4 \pm 3) \times 10^6$. To obtain the final uncertainty we calculated the interval of variation of the mass obtained when using the two extreme values of inclination (i.e., 65° and 70°); this results in an additional $\pm 0.08 \times 10^6 \, M_\odot$ uncertainty that, when added to the $2.9 \times 10^6 \, M_\odot$ previously calculated by partial derivatives, yields a total uncertainty of $3 \times 10^6 \, M_\odot$.

In the literature, the most cited mass for the black hole of NGC 4945 is $\sim 1.4 \times 10^6 \, M_\odot$ inside a radius of 0.7 pc and it was calculated by Greenhill et al. (1997) from the H$_2$O maser that resides in the nucleus of this galaxy. Our measurement of $(4.4 \pm 3) \times 10^6 \, M_\odot$ corresponds to the sum of the black hole mass and the mass of the surrounding material inside $6.5 \pm 3.25$ pc. Subtracting the mass measured by Greenhill et al. from our measurement, we obtain a mass of $\sim 3 \times 10^6 \, M_\odot$ between 0.7 pc and 6.5 pc. This material is expected to fall onto the SMBH in a less than $10^8$ yr according to Storchi-Bergmann & Schnorr-Müller (2019). Reservoirs of similar mass have been found in other active nuclei by several authors (e.g., Combes et al. 2014; Casasola et al. 2015; Schinnerer et al. 2000).

NGC 4945 presents, according to van den Bosch (2016), a stellar radial velocity dispersion of $\log(\sigma) = 2.13$ ($\sim 132$ km s$^{-1}$), which places it significantly below the $M$–$\sigma$ relation for quiescent galaxies of Kormendy & Ho (2013). It is interesting to determine whether the mass needed to reach the relation is available in the proximity of the nucleus. The mass enclosed at each radius of the circumnuclear region can be estimated as above for all resolved radii, and hence we can determine the radius that encloses the mass necessary for the NGC 4945 nucleus to reach the relation assuming that the value of $\sigma$ remains constant during accretion (considering that the dynamical time for the variation in stellar radial velocity dispersion would be greater than the nuclear accretion time, this hypothesis would be reasonable). In order to do this the NGC 4945 SMBH needs to accrete $5.22 \times 10^7 \, M_\odot$ as calculated from the relation and its present position in the $M$–$\sigma$ plane. According to the Keplerian approximation, this amount of mass is enclosed in a 20 pc radius around the center and would be accreted by the black hole in less than $10^8$ yr (e.g., Storchi-Bergmann & Schnorr-Müller 2019), which is coincident with the duty cycle of the AGN (e.g., Emsellem et al. 2015; Storchi-Bergmann & Schnorr-Müller 2019). In this context, the mass necessary for NGC 4945 to reach the $M$–$\sigma$ relation for quiescent galaxies is already located in the immediate vicinity of the nucleus and would be accreted during one duty cycle. Considering that the Eddington luminosity is $L_{\rm Edd} = 1.26 \times 10^{38} \, M_{\rm bh}/M_\odot$ erg s$^{-1}$, the current $L_{\rm Edd}$ is $1.764 \times 10^{44}$ erg s$^{-1}$, while $L_{\rm Edd}$ after all this material falls into the BH would be $L_{\rm Edd} = 6.05 \times 10^{44}$ erg s$^{-1}$. Moreover, assuming that the typical Eddington ratio in Seyfert 2 galaxies is $\log(L/L_{\rm Edd}) = -0.47$ (Bian & Gu 2007), the current bolometric luminosity is $L_{\rm bol} = 6 \times 10^{43}$ erg s$^{-1}$, while that after all the material falls into the BH would be $L_{\rm bol} = 2 \times 10^{44}$ erg s$^{-1}$. The luminosity would increase by a factor of three, consequently increasing the nuclear activity, with an energy output near to the typical bolometric luminosity of a Seyfert nucleus, $L_{\rm bol} \sim 3 \times 10^{44}$ erg s$^{-1}$ (e.g., Singh et al. 2011).

### 3.6. Extended Resolved Dust

Hot dust heated by UV photons originating in an AGN accretion disk can reach temperatures between 800 and 1500 K as measured by different authors (e.g., Glass & Moorwood 1985; Alonso-Herrero et al. 1998; Kishimoto et al. 2011; Burtscher et al. 2015). This places the peak of the emission of this dust in the MIR, around 3–5 $\mu$m. The blue tail of the distribution of this emission will fall in the $K$ band, producing a red excess (e.g., Edelson & Malkan 1986; Alonso-Herrero et al. 1998; Ferruit et al. 2004; Ramos Almeida et al. 2009; Prieto et al. 2010; Schnülle et al. 2013; Durré & Mould 2018; Gravity Collaboration et al. 2020). This phenomenon is ubiquitous in type 1 nuclei, where according to the Unified Model the inner part of the dusty absorber is reached (e.g., Kobayashi et al. 1993), but rarer in type 2 nuclei although there are some examples (e.g., Ferruit et al. 2004; Durré & Mould 2018; Gaspar et al. 2019; Gravity Collaboration et al. 2020). When the red excess is present, the slope of the continuum increases toward red wavelengths, causing a flattening or even a rising of the continuum. In type 2 nuclei, where there is no contribution to the continuum from the AGN accretion disk, the final slope of the continuum will be the sum of the red tail of the stellar emission and the blue tail of the dust emission (Burtscher et al. 2015; Durré & Mould 2018).

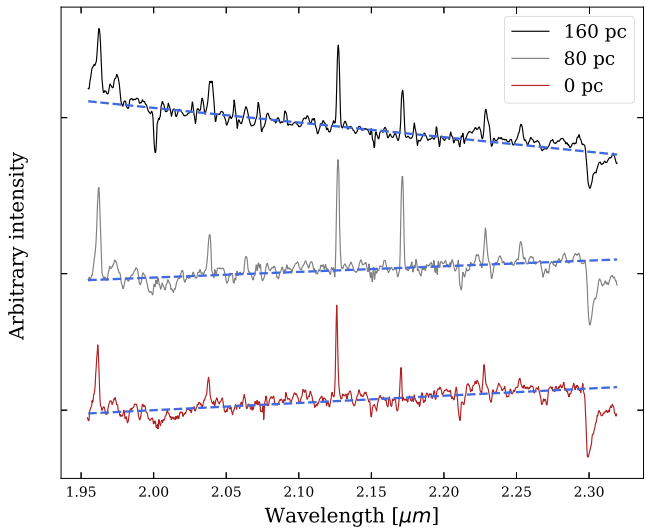

**Figure 9.** Three examples of spectra of NGC 4945: in black, a 20 pc wide extraction centered on 160 pc, where we consider that emission is free from nuclear contributions; in gray, a 6.5 pc wide extraction centered on 80 pc; in red, the same nuclear spectrum as shown in Figure 3. The blue dotted lines highlight the slope of each spectrum, providing evidence of an important change along the displayed range of radius.

In Figure 9 spectra extracted from different spatial regions are shown. Bottom: the nucleus, centered on 0 pc and 3.25 pc wide; center: an intermediate region centered on 80 pc and 6.5 pc wide; top: an external region centered on 160 pc and ∼20 pc wide, which here we assume to be representative of the stellar population with zero contribution from the nuclear emission.

We have found that the slope of the continuum rises slightly toward longer wavelengths in the nuclear spectrum. It is approximately flat for intermediate radii up to 80 pc in the NE direction and up to 65 pc in the SW direction. Beyond that the continuum drops in the red part of the band, leaving only the stellar contribution, which rises toward the blue. The reader can inspect the rich set of spectra in the Appendix, where the gradual changes in emission-line ratios, velocity, and widths are easily spotted.

In order to test whether this effect is caused by the presence of hot dust in the inner 80 pc we built up a stellar template by averaging two 20 pc wide spectra extracted 160 pc on either side of the nucleus, where no contribution from nuclear emission is observed. Previous to the averaging, all spectral features were removed in order to leave only the continuum component. The obtained template was then subtracted from the inner spectra, leaving excesses that can be fitted with blackbody spectra.

All the excess can be fitted by a blackbody function with temperature between 1010 and 1100 K, reinforcing the idea that this emission arises in hot dust located in the circumnuclear region. We have been able to detect hot dust with a spatial sampling of 6.5 pc that forms a large structure of ∼80 pc radius NE and 65 pc SW. In Figure 10 we present three examples of blackbody fitting performed at the nucleus and at 20 and 40 pc in the NE direction. In the figure, the residuals of the fits are presented and blackbody models with different temperatures were added for comparison purposes. The dependence of the temperature measured by this method on the parameters involved is explored in G. Gaspar et al. (2022, in preparation), where Nirdust, a Python package for measuring hot dust temperature in Type 2 AGNs, is presented.

The hot dust with temperatures of 1010–1100 K that we found inside a radius of 80 ± 3.25 pc must be spatially coexistent with the starburst activity found by other authors. However, the UV radiation produced in young OB stars is not enough to heat the dust to these temperatures (e.g Marshall et al. 2018). This can be easily seen in the characteristic spectral energy distribution of AGNs where the emission from dust heated by starburst activity peaks in the MIR (∼60–100 $\mu$m) while the emission from dust heated by an AGN accretion disk peaks at ∼3–5 $\mu$m, with its blue tail penetrating the $K$ band (see, for example, Dwek et al. 2007; Gravity Collaboration et al. 2020). In other words, even though the starburst activity and the hot dust are observed in the same region, the source of photons that heat this dust cannot be young stars. The radiation has to be harder, with the AGN accretion disk being the most probable source. This does not means that it cannot be coexistent with regions of lower-temperature dust heated by the starburst, but that lower-temperature dust would be invisible in the $K$ band.

As mentioned, NGC 4945 is a widely studied galaxy. Other authors have made descriptions of its nucleus in different bands and found evidence of possible nuclear structures with extents comparable to the hot dust found in this work.

Sosa-Brito et al. (2001) presented a $K$-band spectrum taken with the Siding Spring 3.9 m telescope in Australia. Even though the spectrum was extracted with a seeing-limited aperture of 2″.3 (41.4 pc for NGC 4945), the flattening of the continuum slope is evident. In order to detect the dominant contribution to the emission (Seyfert or starburst) they proceeded in an opposite manner to us: they subtracted the 2″.3 aperture spectrum from a wider spectrum extracted in a 4″ (∼72 pc) aperture, both of them centered on the intensity peak. After the subtraction the continuum slope remains unchanged, indicating that the mixing of emission mechanisms for the continuum is the same in the nuclear region and in the circumnuclear region inside 72 pc.

Furthermore, estimations for the far-infrared (100 $\mu$m) luminosity for the NGC 4945 nucleus show that ∼75% of it originates in an elongated region of 222 pc × 167 pc (Brock et al. 1988). This suggests the presence of large amounts of cold dust confined in that region.

Other authors have found starburst emission confined inside a disk or ring on scales ⩽200 pc: Marconi et al. (2000) found a ring mapped in Pa$\alpha$ (1.87 $\mu$m) with a 100 pc size using HST-NICMOS images. Moreover, in an $H - K$ color map they found a structure that resembles a face-on disk with a radius of 4″.5 (∼80 pc). Pérez-Beaupuits et al. (2011) found that the ratio [Ne III] 15.55 $\mu$m/[Ne II] 12.81 $\mu$m, a good tracer of star formation activity, is coincident with the starburst found by Marconi et al. (2000).

Finally, Bendo et al. (2016) found in ALMA data that the continuum at 85.69 GHz and the H42$\alpha$ line originate in a structure that can be modeled as an exponential disk with a scale length radius of ∼2″.1 (∼38 pc). The authors propose that the spatial extent, the absence of broad lines, and the fact that the intensity does not increases toward the center indicate that this emission originates in star formation processes and not in the active nuclei.

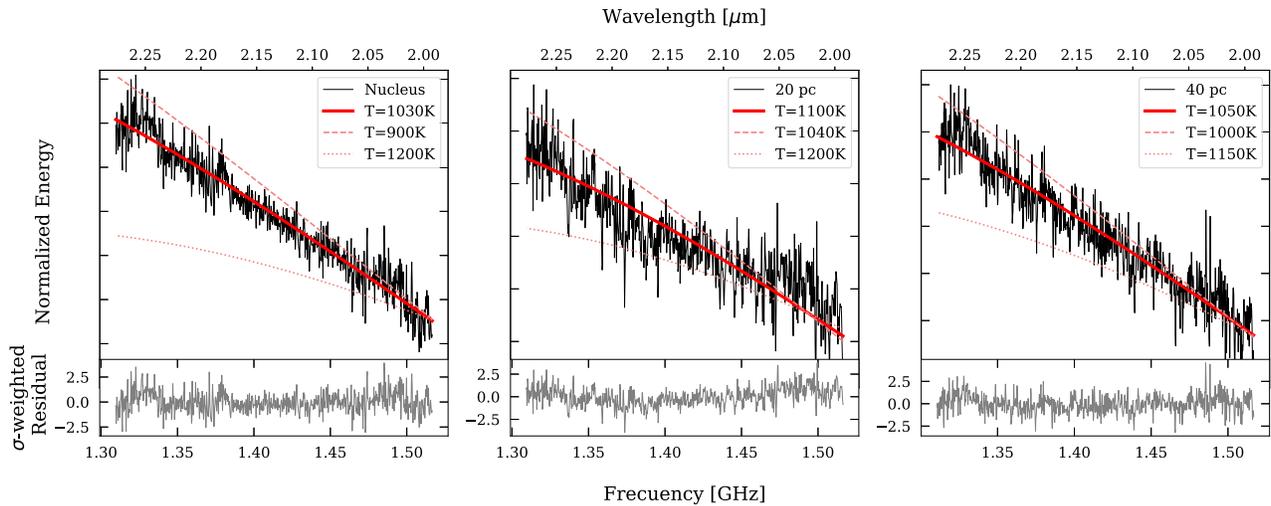

**Figure 10.** Three examples of the blackbody fitting performed over the stellar-subtracted-continuum. The red solid curves correspond to the best fit and the dashed/dotted curves are blackbody functions with different temperatures added for comparison purposes. In the bottom box of each panel the $\sigma$-weighted residual of the fit is presented.

The coexistence of such a variety of emissions in the circumnuclear region of this galaxy is certainly puzzling but points in the direction of the existence of a large (∼80 pc) structure with a density distribution clumpy enough to allow these different dust temperatures to persist.

## 4. Conclusions

NGC 4945 is a widely studied galaxy due to its proximity and the high obscuration that presents in its nuclear region, which has motivated long debates regarding its nuclear activity. In this work we used the infrared spectral range to unveil the structures and physical state of the gas in the circumnuclear region of NGC 4945, using data from F2 and T-ReCS, both from Gemini South. Our main results can be summarized as follows.

1. F2 and T-ReCS images reveal a clumpy spiral structure in a circumnuclear disk that is coincident with the Pa$\beta$ ring-like structure reported by other authors. The detected knots have no optical counterparts and they are probably young star clusters deeply embedded in dust, because they are prominent in the T-ReCS bands. None of these regions are coincident with the supercluster candidates reported by Emig et al. (2020) using ALMA data but the clumps' overall distribution conforms to a circumnuclear disk structure. This disk has a Sérsic ($n = 1$) profile with scale length of 117 pc in the pure continuum emission profile of the F2 spectra at 2 $\mu$m.
2. We report the detection of a bright IR source, SWK, a compact region placed at 55 pc from the *K*-band nucleus in the SW direction along PA = 43°. This source is prominent in the Br$\gamma$ line profile as well as in the T-ReCS bands $Q_a$ and $N$ and presents higher excitation in the H$_2$ diagnostic diagram. The unresolved core has a *K*-band absolute magnitude typical of SSCs found in luminous infrared galaxies in the nearby universe.
3. We provide the first *J*, *H*, and $K_s$ photometric measurements of the nucleus on subarcsecond scales, as well as new accurate values for the aperture surface brightness in the central region of this galaxy. Within the uncertainties of the calibration and the flux measurements available in the literature, we do not detect nuclear variability in the NIR spectral range.
4. The nuclear spectrum of NGC 4945 does not reveal any emission-line AGN feature in the *K*-long band as seen by F2. Moreover, the FWHM/EW ratio for the Br$\gamma$ line presents a significant drop in the nucleus in relation to the circumnuclear regions. These results combined indicate a deeply buried AGN in consonance with previous works.
5. We have found that the slope of the spectral continuum is flattened for spectra extracted inside 80 pc as expected from the presence of hot dust. Using external spectra as a stellar template and a blackbody fitting, we have found temperatures of 1010–1100 K inside 80 pc. The fact that other authors have observed dust at lower temperatures in the same region points to a scenario where the material is clumpy enough to allow the hard AGN photons to heat the dust at long distances while other regions remain shielded and hence colder.
6. From the H$_2$ radial velocity curve, we have estimated the mass contained within a 6.26 pc radius. By subtracting the mass of the central black hole determined via the maser, we report a mass of $M/M_\odot = (3 \times 10^6)$ for the material reservoir available for the dark object's growth within that aforementioned radius. We have also established that the mass necessary for NGC 4945 to reach the $M$–$\sigma$ relation for quiescent galaxies is already located in the immediate vicinity of the nucleus and would be accreted during one duty cycle.

Studies such as the one presented in this paper, making use of instrumental facilities in spectroscopy and imaging in different NIR and MIR spectral bands, are making it possible to describe the circumnuclear regions on scales of less than 100 pc with unprecedented spatial resolution and sensitivity. While many aspects of nuclear fueling dynamics remain unknown, this type of work aims to pave the way for future lines of research on the doorstep of the new generation of 30 m telescopes and the James Webb Space Telescope, which will bring us closer to unraveling the true nature and geometry of these obscured nuclei.

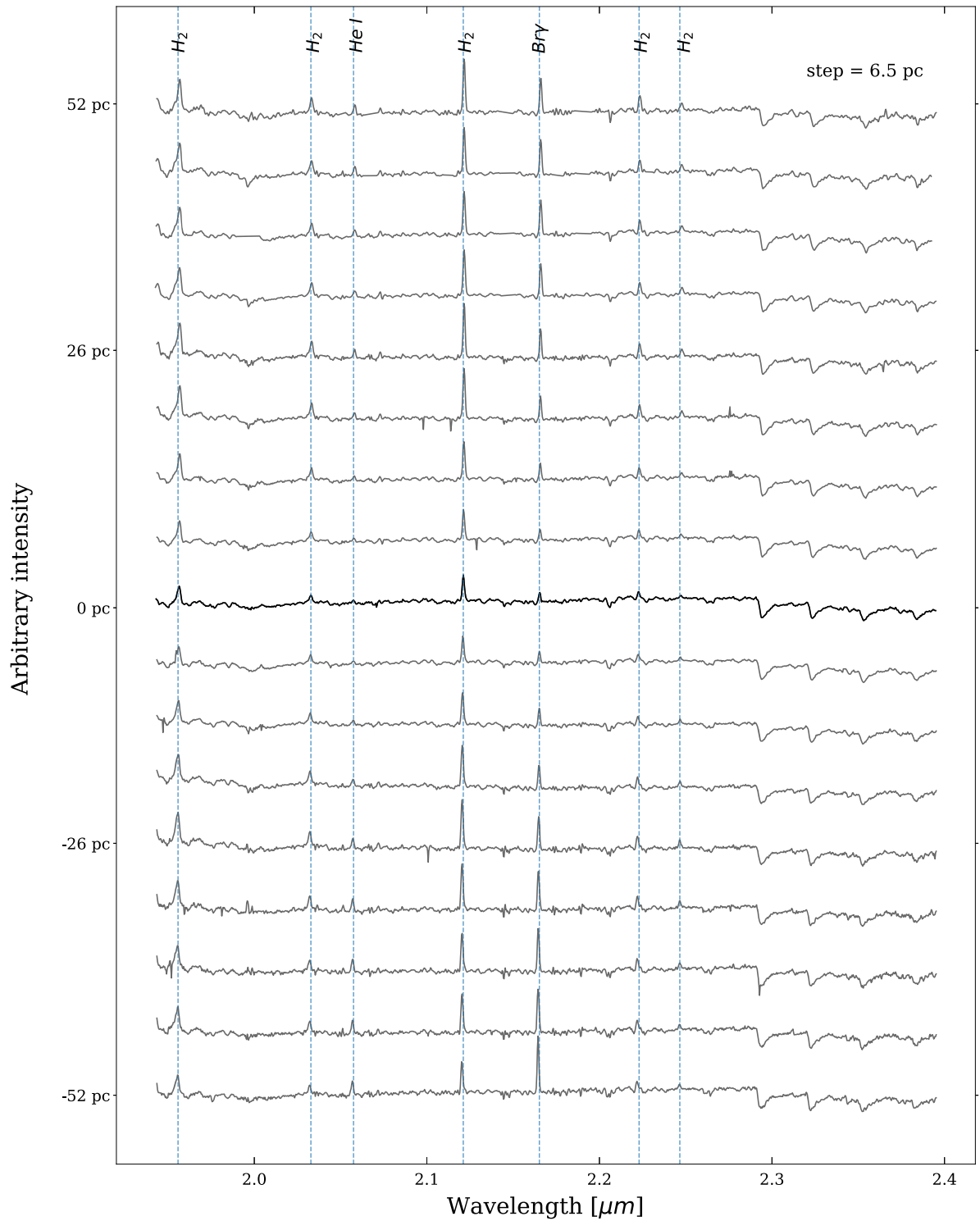

**Figure 11.** Spectra extracted in the region ±52 pc. Negative radius correspond to the SW direction. The step between spectra corresponds to 6.5 pc or equivalently 0″.36. The nuclear spectrum is highlighted in black. It can be noted how the ionized gas lines He I 2.058 $\mu$m and Br$\gamma$ 2.16 $\mu$m increase their intensities in relation to the H$_2$ 2.12 $\mu$m molecular line for increasing radius in both directions while the rest of the H$_2$ lines remain constant. In particular, around −50 pc, where the SWK source is located, this effect is maximum (the effect continues in the next figure).

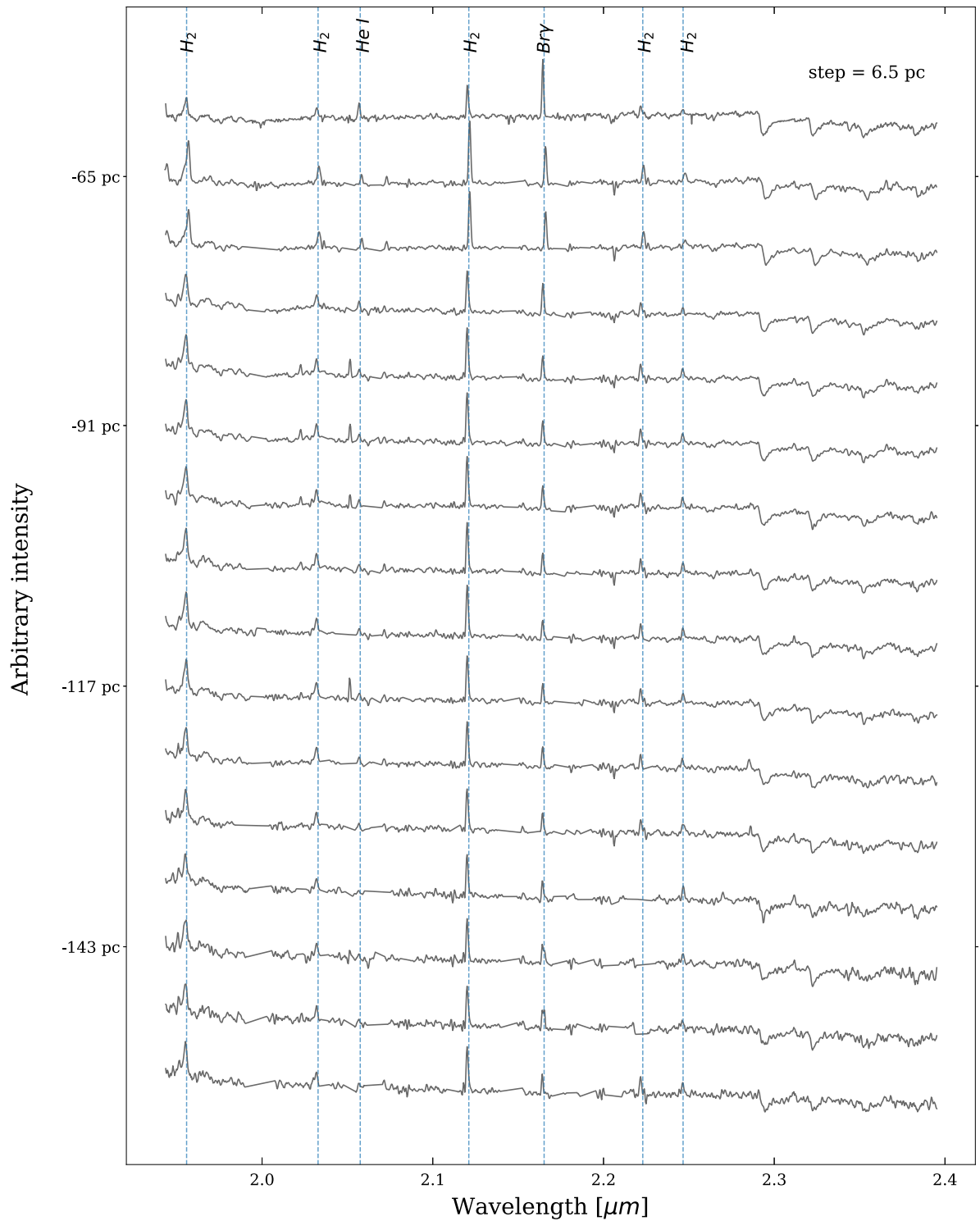

**Figure 12.** Spectra extracted from −58.5 to −156 pc. Negative radius correspond to the SW direction. The step between spectra corresponds to 6.5 pc or equivalently 0″.36. Here the increase in intensity of the ionized gas lines He I 2.058 $\mu$m and Br$\gamma$ 2.16 $\mu$m that started to be noticeable around −30 pc continues up to 65 pc, where it starts to diminish. For larger radii the intensity of these lines remains relatively constant.

Tecnología e Innovación Productiva (Argentina), and Ministério da Ciência, Tecnologia e Inovação (Brazil). This research has made use of the NASA/IPAC Extragalactic Database (NED), which is operated by the Jet Propulsion Laboratory, California Institute of Technology, under contract with the National Aeronautics and Space Administration. Based on observations made with the NASA/ESA Hubble Space Telescope, obtained from the data archive at the Space Telescope Science Institute. STScI is operated by the Association of Universities for Research in Astronomy, Inc. under NASA contract NAS 5-26555. G.Gaspar has a fellowship from Consejo Nacional de Investigaciones Científicas y Técnicas, CONICET, Argentina. This paper has been partially supported with grants from CONICET, Fondo para la

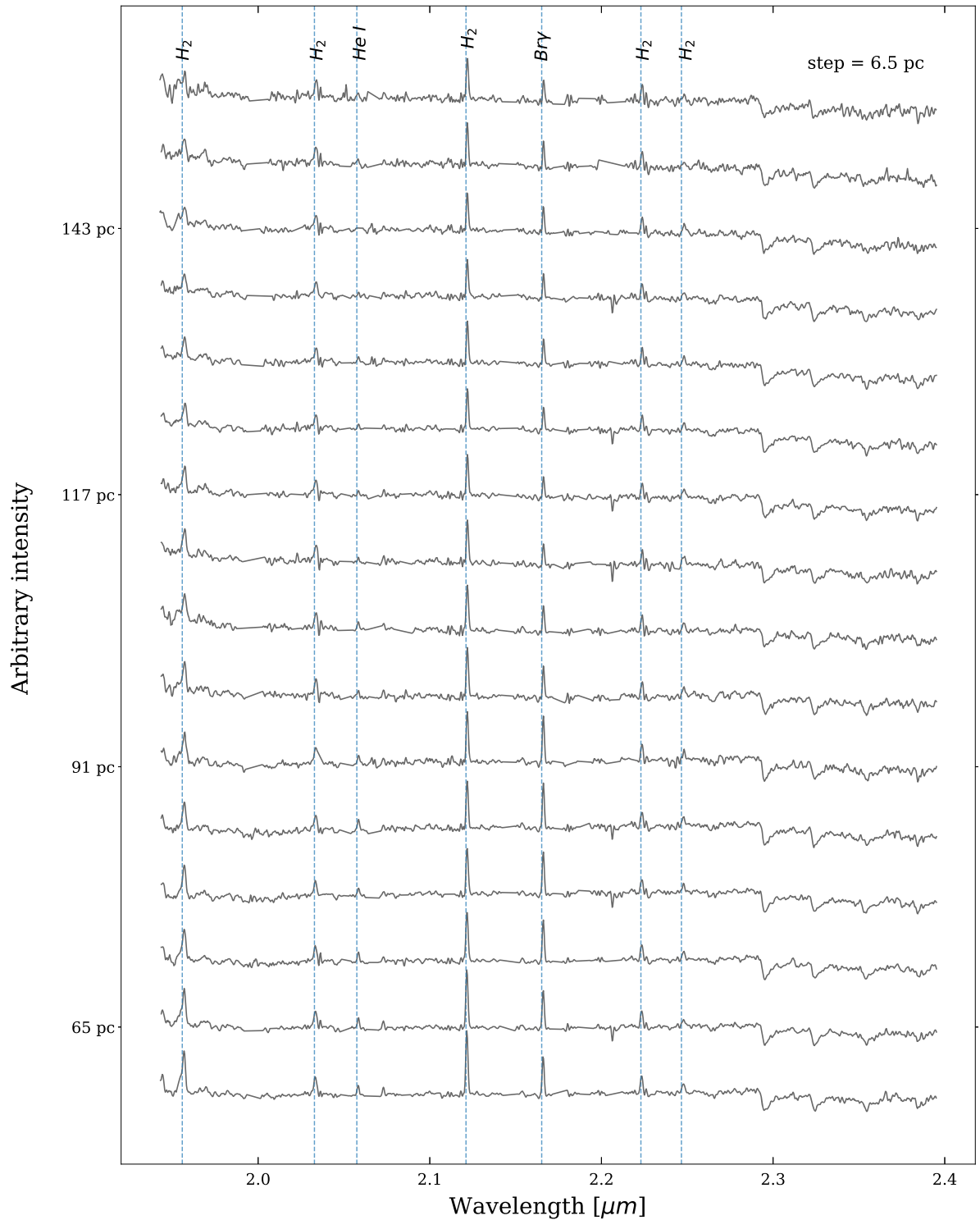

**Figure 13.** Spectra extracted from 58.5 to 156 pc. Positive radius correspond to the NE direction. The step between spectra corresponds to 6.5 pc or equivalently 0″.36. Here the increase in intensity of the ionized gas lines He I 2.058 $\mu$m and Br$\gamma$ 2.16 $\mu$m that started to be noticeable around 26 pc continues up to 91 pc, where it starts to diminish. For larger radii the intensity of these lines remains relatively constant. This could be indicative of a second prominent star formation cluster, but it is not present in either in the F2 profiles/images or the T-ReCS images.

Investigación Cientfica y Tecnológica (FonCyT, PICT-2017-3301), and Secretaría de Ciencia y Tecnología, Universidad Nacional de Córdoba, SeCyT-UNC, Argentina. We thank the anonymous referee for many helpful comments that improved the quality of this paper.

## Appendix

Figures 11, 12, and 13 display all the spectra we extracted for NGC 4945; the step between apertures is 6.5 pc or equivalently 0″.36. Figure 11 displays spectra for the nuclear region up to ∼±50 pc while Figures 12 and 13 display the rest

of the negative and positive radii respectively. The three figures show a change of slope in the continuum: in particular, inside the region delimited by −65 to 90 pc the continuum is flatter with a slight rise in some of the most nuclear spectra. Outside this region the slope of the continuum starts to turn negative with increasing radius. As discussed in Section 3.6 this change in the slope of the continuum can be attributed to hot dust (600–1500 K), and we have proven this scenario by performing blackbody fitting to the stellar-subtracted-continuum up to 65 pc. The captions of the figures provide more details about the displayed spectral features.

## ORCID iDs

G. Gaspar https://orcid.org/0000-0001-9293-4449
R. J. Díaz https://orcid.org/0000-0001-9716-5335
D. Mast https://orcid.org/0000-0003-0469-3193
M. P. Agüero https://orcid.org/0000-0002-4450-5655
M. Schirmer https://orcid.org/0000-0003-2568-9994
G. Günthardt https://orcid.org/0000-0003-3501-5360
E. O. Schmidt https://orcid.org/0000-0001-6048-9715

## References

Aalto, S., Falstad, N., Muller, S., et al. 2020, A&A, 640, A104
Ackermann, M., Ajello, M., Allafort, A., et al. 2012, ApJ, 755, 164
Agüero, M. P., Díaz, R., & Dottori, H. 2016, IJAA, 6, 219
Alonso-Herrero, A., Simpson, C., Ward, M. J., & Wilson, A. S. 1998, ApJ, 495, 196
Antonucci, R. 1993, ARA&A, 31, 473
Bendo, G. J., Henkel, C., D'Cruze, M. J., et al. 2016, MNRAS, 463, 252
Bian, W., & Gu, Q. 2007, ApJ, 657, 159
Blandford, R. D., & Rees, M. J. 1978, PhyS, 17, 265
Brock, D., Joy, M., Lester, D. F., et al. 1988, ApJ, 329, 208
Burtscher, L., Orban de Xivry, G., Davies, R. I., et al. 2015, A&A, 578, A47
Capelo, P. R., & Dotti, M. 2017, MNRAS, 465, 2643
Cardelli, J. A., Clayton, G. C., & Mathis, J. S. 1989, ApJ, 345, 245
Carilli, C. L., Perley, R. A., Dhawan, V., & Perley, D. A. 2019, ApJL, 874, L32
Casasola, V., Hunt, L., Combes, F., & García-Burillo, S. 2015, A&A, 577, A135
Combes, F., García-Burillo, S., Audibert, A., et al. 2019, A&A, 623, A79
Combes, F., García-Burillo, S., Casasola, V., et al. 2013, A&A, 558, A124
Combes, F., García-Burillo, S., Casasola, V., et al. 2014, A&A, 565, A97
Díaz, R., Dottori, H., Vera-Villamizar, N., & Carranza, G. 2003, ApJ, 597, 860
Díaz, R. J., Gómez, P., Schirmer, M., et al. 2013, BAAA, 56, 457
Durré, M., & Mould, J. 2018, ApJ, 867, 149
Dwek, E., Galliano, F., & Jones, A. P. 2007, ApJ, 662, 927
Edelson, R. A., & Malkan, M. A. 1986, ApJ, 308, 59
Eikenberry, S., Bandyopadhyay, R., Bennett, J. G., et al. 2012, Proc. SPIE, 8446, 84460I
Eikenberry, S., Elston, R., Raines, S. N., et al. 2008, Proc. SPIE, 7014, 70140V
Emig, K. L., Bolatto, A. D., Leroy, A. K., et al. 2020, ApJ, 903, 50
Emsellem, E., Renaud, F., Bournaud, F., et al. 2015, MNRAS, 446, 2468
Erben, T., Schirmer, M., Dietrich, J. P., et al. 2005, AN, 326, 432
Falcón-Barroso, J., Bacon, R., Bureau, M., et al. 2006, MNRAS, 369, 529
Falcón-Barroso, J., Ramos Almeida, C., Böker, T., et al. 2014, MNRAS, 438, 329
Ferruit, P., Mundell, C. G., Nagar, N. M., et al. 2004, MNRAS, 352, 1180
Gaia Collaboration, Brown, A. G. A., Vallenari, A., et al. 2018, A&A, 616, A1
Garcia-Burillo, S., Alonso-Herrero, A., Almeida, C. R., et al. 2021, A&A, 652, A98
García-Burillo, S., Combes, F., Usero, A., et al. 2014, A&A, 567, A125
Gaspar, G., Díaz, R. J., Günthardt, G., et al. 2012, BAAA, 55, 297
Gaspar, G., Díaz, R. J., Mast, D., et al. 2019, AJ, 157, 44
Glass, I. S., & Moorwood, A. F. M. 1985, MNRAS, 214, 429
Gomez, P. L., Diaz, R., Pessev, P., et al. 2012, AAS Meeting Abstracts, 219, 413.07
Goulding, A. D., & Alexander, D. M. 2009, MNRAS, 398, 1165
Gravity Collaboration, Pfuhl, O., Davies, R., et al. 2020, A&A, 634, A1
Greenhill, L. J., Moran, J. M., & Herrnstein, J. R. 1997, ApJL, 481, L23
Günthardt, G. I., Agüero, M. P., Camperi, J. A., et al. 2015, AJ, 150, 139
Hönig, S. F., & Kishimoto, M. 2017, ApJL, 838, L20
Kim, W.-T., & Elmegreen, B. G. 2017, ApJL, 841, L4
Kishimoto, M., Hönig, S. F., Antonucci, R., et al. 2011, A&A, 527, A121
Kobayashi, Y., Sato, S., Yamashita, T., Shiba, H., & Takami, H. 1993, ApJ, 404, 94
Koornneef, J. 1993, ApJ, 403, 581
Kormendy, J., & Ho, L. C. 2013, ARA&A, 51, 511
Krolik, J. H., & Begelman, M. C. 1988, ApJ, 329, 702
Madejski, G., Życki, P., Done, C., et al. 2000, ApJL, 535, L87
Malkan, M. A., Gorjian, V., & Tam, R. 1998, ApJS, 117, 25
Marconi, A., Oliva, E., van der Werf, P. P., et al. 2000, A&A, 357, 24
Marshall, J. A., Elitzur, M., Armus, L., Diaz-Santos, T., & Charmandaris, V. 2018, ApJ, 858, 59
Martini, P., Pogge, R. W., Ravindranath, S., & An, J. H. 2001, ApJ, 562, 139
Martini, P., Regan, M. W., Mulchaey, J. S., & Pogge, R. W. 2003, ApJ, 589, 774
Mathis, J. S. 1990, ARA&A, 28, 37
Mazzalay, X., Saglia, R. P., Erwin, P., et al. 2013, MNRAS, 428, 2389
Mingozzi, M., Cresci, G., Venturi, G., et al. 2019, A&A, 622, A146
Moorwood, A. F. M., & Glass, I. S. 1984, A&A, 135, 281
Moorwood, A. F. M., & Oliva, E. 1988, A&A, 203, 278
Moorwood, A. F. M., & Oliva, E. 1990, A&A, 239, 78
Moorwood, A. F. M., van der Werf, P. P., Kotilainen, J. K., Marconi, A., & Oliva, E. 1996, A&A, 308, L1
Nenkova, M., Ivezić, Ž., & Elitzur, M. 2002, ApJL, 570, L9
Nenkova, M., Sirocky, M. M., Ivezić, Ž., & Elitzur, M. 2008, ApJ, 685, 147
Novak, G. S., Ostriker, J. P., & Ciotti, L. 2011, ApJ, 737, 26
Peng, F.-K., Zhang, H.-M., Wang, X.-Y., Wang, J.-F., & Zhi, Q.-J. 2019, ApJ, 884, 91
Pérez-Beaupuits, J. P., Spoon, H. W. W., Spaans, M., & Smith, J. D. 2011, A&A, 533, A56
Pogge, R. W., & Martini, P. 2002, ApJ, 569, 624
Pons, E., McMahon, R. G., Simcoe, R. A., et al. 2019, MNRAS, 484, 5142
Prieto, M. A., Reunanen, J., Tristram, K. R. W., et al. 2010, MNRAS, 402, 724
Ramos Almeida, C., Pérez García, A. M., & Acosta-Pulido, J. A. 2009, ApJ, 694, 1379
Ramos Almeida, C., & Ricci, C. 2017, NatAs, 1, 679
Randriamanakoto, Z., Vaisanen, P., & Escala, A. 2015, IAUGA, 29, 2253176
Reunanen, J., Kotilainen, J. K., & Prieto, M. A. 2002, MNRAS, 331, 154
Riffel, R. A., Storchi-Bergmann, T., Riffel, R., et al. 2021, MNRAS, 504, 3265
Rodríguez-Ardila, A., Pastoriza, M. G., Viegas, S., Sigut, T. A. A., & Pradhan, A. K. 2004, A&A, 425, 457
Rosenthal, M. J., Zaw, I., Greenhill, L., & Zhang, Y. 2020, AAS Meeting Abstracts, 235, 233.01
Schawinski, K., Koss, M., Berney, S., & Sartori, L. F. 2015, MNRAS, 451, 2517
Schinnerer, E., Eckart, A., Tacconi, L. J., Genzel, R., & Downes, D. 2000, ApJ, 533, 850
Schirmer, M. 2013, ApJS, 209, 21
Schmidt, E. O., Mast, D., Díaz, R. J., et al. 2019, AJ, 158, 60
Schnülle, K., Pott, J. U., Rix, H. W., et al. 2013, A&A, 557, L13
Schurch, N. J., Roberts, T. P., & Warwick, R. S. 2002, MNRAS, 335, 241
Simões Lopes, R. D., Storchi-Bergmann, T., de Fátima Saraiva, M., & Martini, P. 2007, ApJ, 655, 718
Singh, V., Shastri, P., & Risaliti, G. 2011, A&A, 533, A128
Skrutskie, M. F., Cutri, R. M., Stiening, R., et al. 2006, AJ, 131, 1163
Sosa-Brito, R. M., Tacconi-Garman, L. E., Lehnert, M. D., & Gallimore, J. F. 2001, ApJS, 136, 61
Spoon, H. W. W., Koornneef, J., Moorwood, A. F. M., Lutz, D., & Tielens, A. G. G. M. 2000, A&A, 357, 898
Storchi-Bergmann, T., McGregor, P. J., Riffel, R. A., et al. 2009, MNRAS, 394, 1148
Storchi-Bergmann, T., & Schnorr-Müller, A. 2019, NatAs, 3, 48
Telesco, C. M., Pina, R. K., Hanna, K. T., et al. 1998, Proc. SPIE, 3354, 534
Tully, R. B., Courtois, H. M., Dolphin, A. E., et al. 2013, AJ, 146, 86
van den Bosch, R. C. E. 2016, ApJ, 831, 134
Venturi, G., Marconi, A., Mingozzi, M., et al. 2017, FrASS, 4, 46